\documentclass[12pt]{article}
\usepackage{amssymb}
\usepackage{amsthm,amssymb,amsmath}
\usepackage[british]{babel}

 1 
1

\newcommand{\be}{\begin{equation}}
\newcommand{\ee}{\end{equation}}
\input{tcilatex}

\begin{document}

\title{\textsc{Nonlinear Dynamics on the Plane and Integrable Hierarchies of
Infinitesimal Deformations }}
\author{B. Konopelchenko\textsc{\thanks{
Permanent address: Dipartimento di Fisica, Universita di Lecce and
Sezione INFN, 73100 Lecce, Italy. B. Konopelchenko is supported in part by
COFIN 2000 ''Sintesi''.}} and L. Mart\'{i}nez Alonso\textsc{\thanks{
L. Martinez Alonso is on leave of absence from Departamento de Fisica
Teorica II, Universidad Complutense, E-28040 Madrid, Spain and is supported
by \ the Fundacion Banco Bilbao Vizcaya Argentaria}} 
\\
\emph{Isaac Newton Institute for Mathematical Sciences,}\\
\emph{University of Cambridge, 20 Clarkson Road, }\\
\emph{Cambridge, CB3 0EH, United Kingdom} }
\date{}
\maketitle

\begin{abstract}
A class of nonlinear problems on the plane, described by nonlinear
inhomogeneous $\bar{\partial}$-equations, is considered. It is shown that
the corresponding dynamics, generated by deformations of inhomogeneous terms
(sources) is described by Hamilton-Jacobi type equations associated with
hierarchies of dispersionless integrable systems. These hierarchies are
constructed by applying the quasiclassical $\bar{\partial}$-dressing method.
\end{abstract}

\vspace*{.5cm}

\begin{center}
\begin{minipage}{12cm}

\end{minipage}
\end{center}

\newpage

\section{Introduction}

Dispersionless integrable equations and hierarchies represent a particular
class of integrable systems with a number of peculiar and remarkable
properties \cite{1}-\cite{13}. They arise in various problems of physics and
mathematics from hydrodynamics and quantum field models to the theory of
conformal mappings (see e.g. \cite{9},\cite{14}-\cite{18}).

Recently it was shown \cite{19}-\cite{21} that the dispersionless integrable
hierarchies are amenable to the quasi-classical $\bar{\partial}$-dressing
method and that they are closely connected with the theory of quasiconformal
mappings on the plane \cite{22}-\cite{24} . In these papers it was
demonstrated that dispersionless integrable hierarchies are associated with
the simple nonlinear $\bar{\partial}$-equation $S_{\bar{z}}=W(z,\bar{z},S_z)$%
.

In the present paper we place this observation in a much wider setting.
Namely, we consider a class of nonlinear problems on the plane which can be
described by equations of the type 
\begin{equation}
S_{\bar{z}}=W(z,\bar{z},S_{z})+h(z,\bar{z}),  \label{1}
\end{equation}
where $z,\bar{z}\in \ensuremath{\mathbb{C}}$ and $S$, $h$ and $W$ are
complex-valued functions. Such equations arise in several problems of
hydrodynamics, electrostatics and quasiconformal mappings. We assume that 
\emph{nonlinearity} $W$ and \emph{source} $h$ are separated. We are looking
for solutions of \eqref{1} in the form $S=S_{0}+\widetilde{S}$, where $S_{0}$
is determined by the source $h$. In this way the construction of solutions $%
S $ of the problem \eqref{1} is nothing but the dressing of the background
solution $S_{0}$ by the use of the quasi-classical $\bar{\partial}$-dressing
method.

Here we will concentrate on the study of properties of infinitesimal
deformations for the problem \eqref{1}. We will show that these
infinitesimal deformations ($\frac{\partial S}{\partial t_n}$) obey
universal (independent of the form of $W$) hierarchies of Hamilton-Jacobi
(H-J) type equations, which in turn give rise to associated hierarchies of
dispersionless integrable systems. To derive these systems we use the
quasi-classical $\bar{\partial}$-dressing method. Particular
parametrizations of variations of sources $h$ lead to different hierarchies
of H-J equations and their associated dispersionless systems. Several
concrete examples, like the dispersionless Kadomtsev-Petviashvili (KP) and
the two-dimensional Toda lattice (2DTL) among others relevant examples, are
considered. Equations arising in different \emph{gauges} are also discussed.

\section{General problem, parametrization of sources and the quasi-classical 
$\bar{\partial}$-dressing method}

So we will consider \emph{abstract} nonlinear systems on the complex plane $%
\ensuremath{\mathbb{C}}$ described by an equation of the form 
\begin{equation}  \label{2.1}
S_{\bar{z}}=W(z,\bar{z},S_z)+h(z,\bar{z}).
\end{equation}
The function $W$ which defines the nonlinearity is assumed to be an analytic
function of $S_z$ (i.e. independent of $\bar{S_z}$). The inhomogeneous term $%
h$ can be treated as an external source for the nonlinear system.

Equations of the form \eqref{2.1} arise in different fields of physics and
mathematics. For example:

\begin{enumerate}
\item  Several problems of the plane motion of fluids \cite{25}, generalized
growth and Hele-Shaw problems \cite{26}.

\item  After differentiation with respect to $z$, equation \eqref{2.1} reads 
\begin{equation}  \label{2.2}
S_{z\bar{z}}=\Big(W(z,\bar{z},S_z)\Big)_z+\rho(z,\bar{z}),
\end{equation}
where $\rho\equiv h_z$ , so that it can be treated as the nonlinear Poisson
equation for the potential $S$. Such equation may arise in some special
types of effective potential models in which $W_z$ and $\rho$ describe an
effective nonlinearity and a given external source, respectively.

\item  Under certain conditions solutions of \eqref{2.1} define
quasiconformal mappings of plane domains (see e.g.\cite{27}-\cite{28}).
Equations of the type \eqref{2.1}arise also in the study of extremal
problems for quasiconformal mappings (see \cite{29}).
\end{enumerate}

Thus, the results obtained for equation \eqref{2.1} may have a wide range of
applications.

In our discussion we will consider the class of nonlinear problems %
\eqref{2.1} in which nonlinearity and sources are \emph{separated}, i.e. we
assume that there is a partition of the complex plane 
\begin{equation}  \label{2.4}
\ensuremath{\mathbb{C}}=G_0\bigcup\widetilde{G},\quad G_0\bigcap\widetilde{G}%
=\varnothing,
\end{equation}
such that 
\begin{equation}
\begin{array}{l}
\label{2.3} h(z,\bar{z})=0,\;\; z\in G_0 \\ 
\\ 
W(z,\bar{z},S_z)=0,\;\; z\in\widetilde{G}.
\end{array}
\end{equation}

We will look for solutions of equation \eqref{2.1} of the form 
\begin{equation}  \label{2.5}
S=S_0+\widetilde{S},
\end{equation}
where $\widetilde{S}$ is bounded on $\ensuremath{\mathbb{C}}$ and 
\begin{equation}
\begin{array}{l}
\label{2.6} S_{0,\bar{z}}=0,\;\; z\in G_0 \\ 
\\ 
\widetilde{S}_{\bar{z}}=0,\;\; z\in\widetilde{G}.
\end{array}
\end{equation}

The representation \eqref{2.5}, \eqref{2.6} imposes no constraint on the
solution of \eqref{2.1}.

In virtue of \eqref{2.3}-\eqref{2.6}, equation \eqref{2.1} is equivalent to
the system 
\begin{equation}
\begin{array}{l}
\label{2.7} S_{0,\bar{z}}=h,\;\; z\in \ensuremath{\mathbb{C}} \\ 
\\ 
\widetilde{S}_{\bar{z}}=W(z,\bar{z}, S_{0,z}+\widetilde{S}_z) ,\;\; z\in%
\ensuremath{\mathbb{C}}.
\end{array}
\end{equation}

Thus, $S_0$ is determined by the source $h$ and can be considered as a
background (seed) solution of \eqref{2.1} corresponding to $W\equiv 0$.
Hence, the whole procedure of construction of solutions of \eqref{2.1} (or
the equivalent system \eqref{2.7}) is the dressing method, a well-known
procedure in the theory of integrable equations (see e.g. \cite{30}-\cite{32}%
). A peculiar feature of the dressing method for \eqref{2.1}is that the
corresponding $\bar{\partial}$-equation is a local nonlinear partial
differential equation. This $\bar{\partial}$-equation can be considered as
the quasi-classical limit of the standard nonlocal $\bar{\partial}$-equation 
\cite{19}-\cite{21}.The method of construction of solutions within this
quasiclassical $\bar{\partial}$-dressing method \cite{19}-\cite{21}
consists, basically, in solving \eqref{2.7} or the quasi-linear equation for 
$m\equiv S_z$ in the domain $G_0$ and the gluing with $S$ in the domain $%
\widetilde{G}$ \cite{20}.

In this paper we will concentrate on the study of the properties of
deformations of solutions of equation \eqref{2.1}. Infinitesimal
deformations (variations) $S\rightarrow S+\delta S$, generated by
infinitesimal deformations $h\rightarrow h+\delta h$ of the source, are
determined by the linear inhomogeneous Beltrami equation 
\begin{equation}  \label{2.9}
\Big(\delta S \Big)_{\bar{z}}=W^{\prime}\Big(z,\bar{z},S_z\Big)\Big(\delta S%
\Big)_z+\delta h,
\end{equation}
where $W^{\prime}(z,\bar{z},\xi)\equiv \frac{\partial W(z,\bar{z},\xi)}{%
\partial \xi}$. Properties of the Beltrami equation are well-studied (see
e.g.\cite{33},\cite{22},\cite{23}). For our analysis we need two of them 
\cite{33}:

\begin{enumerate}
\item  If $f_1,\ldots,f_n$ are solutions of the homogeneous Beltrami
equation $f_{\bar{z}}=\mu f_z$, then any arbitrary differentiable function $%
F(f_1,\ldots,f_n)$ is a solution too

\item  If the function $\mu$ satisfies $|\mu|\leq k<1$, then the only
solution $f_{\bar{z}}=\mu f_z$ such that $f_{\bar{z}}$ is locally $L^p$ for
some $p>2$, and such that $f$ vanishes at some point of the extended plane $%
\ensuremath{\mathbb{C}}^{*}$ is $f\equiv 0$.
\end{enumerate}

These two properties provide a basis for the quasiclassical $\bar{\partial}$%
-dressing method \cite{19}, \cite{21}. The first one implies that, together
with the infinitesimal deformations $\delta _{1}S,\delta _{2}S,\ldots
,\delta _{n}S$, any differentiable function of them

$F(\delta _{1}S,\delta _{2}S,\ldots ,\delta _{n}S)$ is also a solution of
the Beltrami equation $f_{\bar{z}}=W^{\prime }f_{z}$ in the domain $G_{0}$.
Furthermore, under the conditions of the second property, if one is able to
find a function $F$ such that $F(\delta _{1}S,\delta _{2}S,\ldots ,\delta
_{n}S)$ is bounded on $\ensuremath{\mathbb{C}}$ and vanishes at some point
of $\ensuremath{\mathbb{C}}^{\ast }$, one has 
\begin{equation}
F(\delta _{1}S,\delta _{2}S,\ldots ,\delta _{n}S)=0.  \label{2.10}
\end{equation}
The derivation of equations of the type \eqref{2.10} is one of the main
goals of the quasiclassical $\bar{\partial}$-dressing method. In this paper
we aim to determine these kind of relations for the infinitesimal
deformations of the problem \eqref{2.1}. To do that one has to parametrize
the source $h$ and its deformations in one or another way.

Having in mind the relation $S_{0,z\bar{z}}=h_{z}=\rho (z,\bar{z})$, it is
quite natural to choose $\rho $ in the form 
\begin{equation}
\rho =h_{z}=\sum_{\alpha =1}^{N}\sum_{n=0}^{\infty }\gamma _{\alpha n}\delta
^{(n,0)}(z-z_{\alpha }),  \label{2.11}
\end{equation}
where $\gamma _{\alpha n}$ and $z_{\alpha }$ are arbitrary complex constants
($z_{\alpha }\in \widetilde{G}$) and $\delta ^{(n,0)}(\xi )$ are the $z$%
-derivatives of Dirac delta functions. Since 
\[
S_{0}(z,\bar{z})=\int \int_{\widetilde{G}}dz^{\prime }\wedge d\bar{z}%
^{\prime }\ln (z-z^{\prime })\,\rho (z^{\prime },\bar{z}^{\prime }), 
\]
one has 
\begin{equation}
S_{0}(z,\bar{z})=\sum_{\alpha =1}^{N}\Big(t_{0\alpha }\ln (z-z_{\alpha
})+\sum_{n=1}^{\infty }\frac{t_{n\alpha }}{(z-z_{\alpha })^{n}}\Big),
\label{2.12}
\end{equation}
where $t_{0\alpha }=\gamma _{\alpha 0}$ and $t_{n\alpha
}=(-1)^{n}(n-1)!\gamma _{n\alpha },\;\;n\geq 1$.

Of course, all pole-type singularities in \eqref{2.12} can be obtained by
coalescing logarithmic terms, so that only the $\delta^{(n,0)}$ terms in %
\eqref{2.11} and the $t_{0\alpha}\ln(z-z_{\alpha})$ terms in \eqref{2.12}
are, in fact, of fundamental importance. But it is convenient to add other
terms to the source $h$ from the very beginning instead of performing
coalescence at the end.

By considering an infinite number of points $z_{\alpha}$ and by passing from
the sum in \eqref{2.11} to an integral, one gets a source $\rho$ of generic
form. However, the simpler forms \eqref{2.11} and \eqref{2.12} are much more
convenient for performing calculations. The symbols $t_{n\alpha}$ and $%
z_{\alpha}$ ($\alpha=1,\ldots,N;\,n\geq 0$) are free parameters, so that
within the class of sources given by \eqref{2.11} the deformations
(variations) of the sources are generated by variations of $t_{n\alpha}$ and 
$z_{\alpha}$. In this case $\delta h=\epsilon\frac{\partial h}{\partial \tau}
$ and $\delta S=\epsilon\frac{\partial S}{\partial \tau}$, where $\epsilon$
is an infinitesimal parameter and $\tau$ is any of the parameters $%
t_{n\alpha}$ and $z_{\alpha}$. Hence, we have 
\begin{equation}  \label{2,13}
(S_{\tau})_{\bar{z}}=W^{\prime}\cdot(S_{\tau})_z+h_{\tau}.
\end{equation}
In virtue of \eqref{2.5} and \eqref{2.12} one has 
\begin{equation}  \label{2.14}
\begin{array}{l}
S_{t_{0\alpha}}=\ln(z-z_{{\alpha}})+\widetilde{S}_{t_{0\alpha}}, \\ 
\\ 
S_{t_{n\alpha}}=\frac{1}{z-z_{{\alpha}}}+\widetilde{S}_{t_{n\alpha}},
\end{array}
\end{equation}
and 
\begin{equation}  \label{2.15}
S_{z_{\alpha}}=-\frac{t_{0\alpha}}{z-z_{{\alpha}}}+ \sum_{n\geq 1} \frac{%
nt_{n\alpha}}{(z-z_{{\alpha}})^{n+1}}+\widetilde{S}_{z_{\alpha}}.
\end{equation}
These expressions have singularities, but all of them are located in the
domain $\widetilde{G}$.

Now let us consider a differentiable function of the type $%
F(S_{t_{n\alpha}},S_{z_{\alpha}},\ldots)$. It is a regular function on the
domain $G_0$. Hence due to the first property of the Beltrami equation, one
has 
\begin{equation}  \label{2.16}
F_{\bar{z}}=W^{\prime}F_z,\quad z\in G_0.
\end{equation}
In general this function has singularities in the domain $\widetilde{G}$ due
to the singular terms of $S_{0\tau}$. But if we manage to cancel these
singularities by a right choice of $F$, then we have 
\begin{equation}  \label{2.17}
F_{\bar{z}}=0,\quad z\in\widetilde{G},
\end{equation}
so that 
\begin{equation}  \label{2.18}
F_{\bar{z}}=W^{\prime}F_z,\quad z\in \ensuremath{\mathbb{C}}.
\end{equation}
In this case, as we are assuming that $\widetilde{S}$ is bounded in $%
\ensuremath{\mathbb{C}}$, if $F(S_{t_{n\alpha}},S_{z_{\alpha}},\ldots)$
vanish at some point of the extended complex plane then from the second
basic property of Beltrami equation we conclude 
\begin{equation}  \label{2.19}
F(S_{t_{n\alpha}},S_{z_{\alpha}},\ldots)=0.
\end{equation}
Equations like \eqref{2.19} are our desired equations relating infinitesimal
symmetries of the system \eqref{2.1}. The next sections of the present paper
are devoted to the derivation of hierarchies of equations of this type. As
we shall see the form of the equations \eqref{2.19} does not depend on the
choice of $W$, so that they have a universal character.

The problem \eqref{2.1} and the equations \eqref{2.19} have an additional
important property; namely, they are invariant under the \emph{gauge
transformations} 
\[
S(z,\bar{z},\tau)\rightarrow S^{\prime}(z,\bar{z},\tau)=S(z,\bar{z}%
,\tau)+g(\tau), 
\]
where $g(\tau)$ is an arbitrary function. For concrete equations this
property has been observed earlier in \cite{10}. This invariance implies
that we can arbitrarily prescribe the value of $S$ at any given point $z_0$
(except at one of the $z_{\alpha}$), i.e. we can always \emph{normalize} $S$
by the condition $S(z_0,\tau)=g_0(t_{n\alpha},z_{\alpha})$, in particular $%
S(z_0)=0$. At the point $z=z_{\alpha}$, one can choose the \emph{gauge} $%
\widetilde{S}(z_{\alpha})=0$. As we shall see different gauges produce
different, but related, equations of the type \eqref{2.19}. It is also
possible to give a gauge invariant formulation of equations \eqref{2.19}.

Before proceeding to the construction of concrete equations we would like to
note the following: \vspace{0.2cm}

\noindent\ \ \ \textbf{Remark 2.1} Equation (3) is the Euler-Lagrange
equation with the Lagrangian

\[
L=\frac{1}{2}S_{z}S_{\overline{z}}-\Phi (z,\overline{z},S_{z})+\rho S 
\]
where $\frac{\partial \Phi (z,\overline{z,}S_{z})}{\partial (S_{z})}=W(z,%
\overline{z},S_{z}).$ Equation (3) can be also represented in the form of
the conservation law

\[
(\frac{1}{2}S_{z})_{\overline{z}}+(\frac{1}{2}S_{\overline{z}}-W(z,\overline{%
z},S_{z})-h)_{z}=0. 
\]
Formally this conservation law is connected with the gauge invariance
mentioned above.

\textbf{Remark 2.2 \ }It is well-known that for the plane potential flows
the kinetic energy of a fluid coincides with the Dirichlet integral (see
e.g.[34]). In our case the total ''kinetic energy'' $\frac{1}{2}\int \int_{%
\Bbb{C}}S_{z}S_{\overline{z}}dz\wedge d\overline{z}$ \ diverges due to
singularity of $S_{0}.$ A regularized ''kinetic energy'' can be defined as
the complex Dirichlet type integral 
\[
D=\frac{1}{2}\int \int_{\Bbb{C}}(\partial S\wedge \overline{\partial }%
S-\partial S_{0}\wedge \overline{\partial }S_{0})=\frac{1}{4i}\int \int_{%
\Bbb{C}}(dS\wedge \ast dS-dS_{0}\wedge \ast dS_{0}).
\]
$\ \ $

\ \bigskip \textbf{Remark 2.3} Within the fluid mechanical interpretation of
the problem \eqref{2.1} the $\gamma _{0\alpha }\delta ^{(0,0)}(z-z_{\alpha
}) $ terms in the source $\rho $ or the logarithmic terms $t_{0\alpha }\ln
(z-z_{\alpha })$ describe sources or sinks of the fluid,while other terms
describe vortices located at the points $z_{\alpha }$ (see e.g. \cite{34}).
In the electrostatic applications the $\gamma _{0\alpha }\delta
^{(0,0)}(z-z_{\alpha })$ terms or $t_{0\alpha }\ln (z-z_{\alpha })$ terms,
obviously correspond to point charged particles with charges $t_{0\alpha }$.
Other terms describe contributions from electrical multipoles. Thus, the
deformations we are discussing here are generated by the infinitesimal
variations of the strengths and positions of sources or sinks of fluid,
charges of the point particles and their positions, and strengths and
positions of vortices and multipoles. \vspace{0.1cm}

\noindent\ \ \ \textbf{Remark 2.4} Equations of the type \eqref{2.19} can be
treated as the dispersionless limit of certain dispersive integrable
hierarchies. Under this limit wave functions $\psi $ of the fully dispersive
case become $\psi =\exp (\frac{S}{\epsilon })$ where $\epsilon \rightarrow 0$%
. Within such a connection the variables $t_{n\alpha }$ represent \emph{slow
variables} associated with the standard flow variables $x_{n\alpha }=\frac{%
t_{n\alpha }}{\epsilon }$. The variables $t_{0\alpha }$ can be considered as
a quasi-classical limit of the Miwa variables. Indeed, in terms of Miwa
variables, the undressed dispersive wave function $\psi _{0}$ has the form $%
\psi _{0}(z)=\prod_{\alpha =1}^{N}(\frac{z}{z_{\alpha }}-1)^{p_{\alpha }}$
where $p_{\alpha }$ are integers (see \cite{35}). By considering large $%
p_{\alpha }$ and by introducing $t_{0\alpha }$ via $p_{\alpha }=t_{0\alpha
}/\epsilon $ ($\epsilon \rightarrow 0$), one gets 
\[
\psi _{0}\sim \exp \frac{\sum_{\alpha =1}^{N}t_{0\alpha }\ln (z-z_{\alpha })%
}{\epsilon }\sim \exp (\frac{S_{0}}{\epsilon }). 
\]

\section{One point-like source case}

We begin by the simplest case of source $\rho$ concentrated on a single
point, i.e. 
\begin{equation}  \label{3.1}
S_0=t_0\ln(z-z_{\alpha})+\sum_{n=1}^{\infty}\frac{t_n}{(z-z_{\alpha})^n}.
\end{equation}
Therefore 
\begin{equation}  \label{3.2}
\begin{array}{l}
\frac{\partial S}{\partial t_0}=\ln(z-z_{\alpha})+\frac{\partial \widetilde{S%
}}{\partial t_0}, \\ 
\\ 
\frac{\partial S}{\partial t_n}=\frac{1}{(z-z_{\alpha})^n}+\frac{\partial 
\widetilde{S}}{\partial t_n}.
\end{array}
\end{equation}
From these equations we deduce 
\begin{equation}  \label{3.3}
e^{-\frac{\partial S}{\partial t_0}}=\frac{1}{z-z_{\alpha}}e^{-\frac{%
\partial \widetilde{S}}{\partial t_0}},\quad z\in\ensuremath{\mathbb{C}},
\end{equation}
and 
\begin{equation}  \label{3.4}
e^{-\mathcal{D}(z_1)S(z)}=\frac{z-z_1}{z-z_{\alpha}}e^{-\mathcal{D}(z_1) 
\widetilde{S}(z)} , \quad z\in\ensuremath{\mathbb{C}},\; z_1\in \widetilde{G}%
,
\end{equation}
where $\mathcal{D}(z)$ stands for the \emph{quasiclassical vertex operator} 
\begin{equation}  \label{3.5}
\mathcal{D}(z):=\sum_{n=1}^{\infty}\frac{1}{n}(z-z_{\alpha})^n \frac{\partial%
}{\partial t_n}.
\end{equation}

Let us first consider variations of the the $t_n$ variables (\emph{time
variables}). By counting singularities of the derivatives $\frac{\partial S}{%
\partial t_n}$, one concludes that the simplest of the equations of the type %
\eqref{2.14} is of the form 
\begin{equation}  \label{3.5}
\begin{array}{l}
S_{t_2}-(S_{t_1})^2-vS_{t_1}-u=0, \\ 
\\ 
v=-2\widetilde{S}_{t_1}(z_{\alpha}),\quad
u=S_{t_2}(z_0)-(S_{t_1}(z_0))^2-vS_{t_1}(z_0),
\end{array}
\end{equation}
where $z_0$ is an arbitrary point $z_0\neq z_{\alpha}$. Other analogous
equations look like 
\begin{equation}  \label{3.6}
S_{t_n}-(S_{t_1})^n-\sum_{k\geq 0}^{n-1}v(S_{t_1})^k=0,\quad n\geq 2,
\end{equation}
where the functions $v_k=v_k(t)$ are appropriate combinations of derivatives
of $\widetilde{S}(z_{\alpha})$ and $S(z_0)$ with respect to $t_1$.

Equations \eqref{3.6} form a complete set of equations for infinitesimal
variations $S_{t_n}$ of $S$, generated by variations of the times $t_n$.
They are equations of Hamilton-Jacobi type.

An important property of equations \eqref{3.6} is that they have the same
form for all functions $W(z,\bar{z},S_z)$ describing nonlinearities in the
problems \eqref{2.1}.

\vspace{0.2cm} \noindent\textbf{Remark 3.1} Within the quasi-classical $\bar{%
\partial}$-dressing method, the functions $W$ play a role of quasi-classical 
$\bar{\partial}$-data \cite{19},\cite{21}. The independence of the form of
integrable equations on the $\bar{\partial}$-data is a general feature of
the dressing method.

By construction, all the equations \eqref{3.6} are compatible and solvable
by the use of the $\bar{\partial}$-problem \eqref{2.1} and, in turn, they
determine a hierarchy of equations for the functions $u,v$ and so on. The
form of these equations depends on the gauge choice.

There are two natural gauges. The first one is $S(z_0)=0,\; (z_0\neq
z_{\alpha})$, in which $u=0$ and the hierarchy \eqref{3.6} becomes the
Hamilton-Jacobi system of equations for the dispersionless modified KP
(dmKP) hierarchy. The lowest member of which is \cite{36}, \cite{21} 
\begin{equation}  \label{3.7}
v_{t_3}+\frac{3}{2}v^2v_{t_1}-\frac{3}{4}v_{t_1}\partial_{t_1}^{-1}(v_{t_2})
-\frac{3}{4}\partial_{t_1}^{-1}(v_{t_2t_2})=0.
\end{equation}

The second natural gauge is $\widetilde{S}(z_{\alpha})=0$. It leads to $v=0$%
, and then \eqref{3.6} reduce to the system of Hamilton-Jacobi equations for
the well-known dKP hierarchy (see e.g.\cite{1}-\cite{12}). The dKP equation
itself is 
\begin{equation}  \label{3.8}
u_{t_3}=\frac{3}{2}uu_{t_1}+\frac{3}{4}\partial_{t_1}^{-1}(u_{t_2t_2}).
\end{equation}

Other gauges can be of interest too. For example, if we impose $S(z_0)=\frac{%
t_1}{z_0-z_{\alpha}}$, i.e. $\widetilde{S}(z_0)=0$, we get a hierarchy of $%
2+1$-dimensional Gardner (mixed KP-mKP) equations. We shall refer to the
hierarchy of equations \eqref{3.6} and associated hierarchy for $u$ and $v$
as the d(KP-mKP) hierarchy.

One can formulate equations in gauge invariant form. In such a formulation
one has equations \eqref{3.6} for the gauge invariant object $%
S^*:=S(z)-S(z_0)$, while equations for \emph{potentials} $v_k$ are
formulated in terms of the gauge-invariant 
\[
w:=u+\frac{1}{2}\partial_t^{-1}v_{t_2}-\frac{1}{4}v^2. 
\]
This formulation provides us also with the dispersionless Miura
transformation \cite{36} 
\[
u=\frac{1}{2}\partial_t^{-1}v_{t_2}-\frac{1}{4}v^2. 
\]

By proceeding along the same lines as in \cite{37}, one can derive several
important generating equations and addition formulae associated with
equations \eqref{3.6}. In virtue of \eqref{3.4}, both $p:=\frac{\partial S}{%
\partial t_1}$ and $\exp(-\mathcal{D}(z_1)S(z))$ have a simple pole at $%
z=z_{\alpha}$, so that one gets 
\begin{equation}  \label{3.9}
p(z)-p(z_0)=-\frac{1}{z_1-z_{\alpha}}\Big(e^{-\mathcal{D}(z_1)S(z)}-e^{-%
\mathcal{D}(z_1)S(z_0)}\Big),
\end{equation}
where $z_0\in\ensuremath{\mathbb{C}}$ is an arbitrary point $z_0\neq
z_{\alpha}$.

Equation \eqref{3.9} is of fundamental importance for describing
infinitesimal deformations. It is the generating equation for the hierarchy
of equations \eqref{3.6}. Indeed, by expanding both sides of \eqref{3.9} in
series on $z_{1}$, one gets a system which is equivalent to \eqref{3.6}.
Furthermore,by evaluating both sides of \eqref{3.9} at $z_{0}=z_{1}$, one
gets 
\begin{equation}
p(z)-p(z_{1})=-\frac{1}{z_{1}-z_{\alpha }}\exp {\Big(-\mathcal{D}%
(z_{1})(S(z)-\widetilde{S}(z_{\alpha }))\Big)},\quad z\in %
\ensuremath{\mathbb{C}},\;z_{1}\in \widetilde{G}.  \label{3.10}
\end{equation}
Thus, by setting $z=z_{2}\in \widetilde{G}$ and using \eqref{3.4} and the
skew-symmetry of the left-hand side of \eqref{3.10} under the interchange $%
z_{1}\leftrightarrow z_{2}$, one concludes that 
\begin{equation}
\mathcal{D}(z_{1})(\widetilde{S}(z_{2})-\widetilde{S}(z_{\alpha }))=\mathcal{%
D}(z_{2})(\widetilde{S}(z_{1})-\widetilde{S}(z_{\alpha })).  \label{3.11}
\end{equation}
Equation \eqref{3.11} implies that there exists a function $F$ such that 
\begin{equation}
\widetilde{S}(z)=\widetilde{S}(z_{\alpha })-\mathcal{D}(z)F.  \label{3.12}
\end{equation}
So one has 
\begin{equation}
p(z_{2})-p(z_{1})=\frac{(z_{2}-z_{1})}{(z_{\alpha }-z_{1})(z_{2}-z_{\alpha })%
}\;e^{\mathcal{D}(z_{1})\mathcal{D}(z_{2})F},\;\;z_{1},z_{2}\in \widehat{G}.
\label{3.13}
\end{equation}
Further, by summing up the relation \eqref{3.13} for pairs of points $%
(z_{2},z_{1}),\;(z_{1},z_{3})$ and $(z_{3},z_{2})$, one gets 
\begin{equation}
\begin{gathered}
(z_1-z_2)(z_3-z_{\alpha})e^{\mathcal{D}(z_1)\mathcal{D}(z_2)F}+
(z_3-z_1)(z_2-z_{\alpha})e^{\mathcal{D}(z_3)\mathcal{D}(z_1)F}\\\\
+(z_2-z_3)(z_1-z_{\alpha})e^{\mathcal{D}(z_2)\mathcal{D}(z_3)F}=0,\quad
z_1,z_2,z_3\in\widetilde{G}. \end{gathered}  \label{3.14}
\end{equation}
It is the addition formula for the dKP hierarchy in the case when the
singularity is located at a finite point $z_{\alpha }$. One can get the
usual form of the dKP addition formula (see e.g.\cite{11},\cite{12},\cite{18}%
) by sending $z_{\alpha }\rightarrow \infty $. Equation \eqref{3.14}, which
is gauge invariant, means that $F=\ln \tau _{dKP}$, where $\tau _{dKP}$
stands for the tau-function of the dKP hierarchy \cite{7},\cite{8},\cite{11}%
, \cite{12}.

From \eqref{3.9} one finds also that 
\begin{equation}
\begin{gathered}
\ln[p(z)-p(z_0)]-\ln(z_{\alpha}-z_i)-\mathcal{D}(z_i)\widetilde{S}(z_{%
\alpha})\\\\ =\ln\Big[
e^{-\mathcal{D}(z_i)S(z)}-e^{-\mathcal{D}(z_i)S(z_0)}\Big],\quad
z_i\in\widetilde{G},\;(i=1,2,3). \end{gathered}  \label{3.15}
\end{equation}
Acting on this equation by the operators $\mathcal{D}(z_{k})$ and $\mathcal{D%
}(z_{l})$ with different values of the indexes $k,l,i$, and summing up the
corresponding equations, one gets 
\begin{equation}
\sum_{i,k,l}\epsilon _{ikl}(\mathcal{D}(z_{i})-\mathcal{D}(z_{k}))\ln \Big[%
e^{-\mathcal{D}(z_{l})S(z)}-e^{-\mathcal{D}(z_{l})S(z_{0})}\Big]=0,\quad
z\in \ensuremath{\mathbb{C}},\;z_{i}\in \widetilde{G},  \label{3.16}
\end{equation}
where $\epsilon _{ikl}$ is the totally antisymmetric tensor ($\epsilon
_{123}=1$). Equation \eqref{3.16} is the generating equation for the
hierarchy of equations for $S(z,\bar{z},{\boldsymbol{t}})$. Indeed, the
taylor expansion of its left-hand side leads to an infinite family of
equations for $S$ only. For example, in the gauge $\widetilde{S}(z_{\alpha
})=0$, the first nontrivial order gives 
\begin{equation}
S_{t_{1}t_{3}}=\frac{3}{4}S_{t_{2}t_{2}}+\frac{3}{2}\Big[%
S_{t_{2}}-(S_{t_{1}})^{2}\Big]S_{t_{1}t_{1}},  \label{3.17}
\end{equation}
that is the $S$-equation for the dKP equation \cite{21}. In the gauge $%
S(z_{0})=0$ $(z_{0}\neq z_{\alpha })$, one has the corresponding equation
for the dmKP hierarchy. Equation \eqref{3.16} encodes all relations between
infinitesimal deformations generated by variations of times $t_{n}$.

Now let us include also into consideration the variations of strength of the
logarithmic term in \eqref{3.1}. Using \eqref{3.2} and \eqref{3.3}, one
readily gets 
\begin{equation}  \label{3.18}
\begin{array}{l}
S_{t_1}=A_{11}e^{-S_{t_0}}+A_{10}, \\ 
\\ 
S_{t_2}=A_{11}^2e^{-2S_{t_0}}+A_{21}e^{-S_{t_0}}+A_{20},
\end{array}
\end{equation}
where 
\begin{equation}  \label{3.20}
\begin{array}{l}
A_{11}=e^{\widetilde{S}_{t_0}(z_{\alpha})},\quad
A_{20}=S_{t_2}(z_0)-A_{11}^2 e^{-2S_{t_0}(z_0)}-A_{21}e^{-S_{t_0}(z_0)} , \\ 
\\ 
A_{21}=2A_{11}\Big(\frac{\partial \widetilde{S}_{t_0}(z)}{\partial z}\Big)%
_{z=z_{\alpha}},\quad A_{10}=S_{t_1}(z_0)-A_{11}e^{-S_{t_0}(z_0)}.
\end{array}
\end{equation}
In general, 
\begin{equation}  \label{3.21}
S_{t_n}=\sum_{k=0}^n A_{nk}e^{-k S_{t_0}},
\end{equation}
where $A_{nk}(t_0,\vec{t})$ are certain functions.

The hierarchy of Hamilton-Jacobi equations \eqref{3.21} is a basic one for
the one-point-like source \eqref{3.1}. Equations \eqref{3.21} imply a
hierarchy of integrable equations for the coefficients $A_{nk}$. The lowest
member of this hierarchy arises from the system of equations \eqref{3.18}
and in the gauge $S(z_0)=0$ looks like 
\begin{equation}  \label{3.22}
\begin{array}{l}
\phi_{t_1}+\psi_{t_0}-e^{\phi_{t_0}}=0, \\ 
\\ 
\phi_{t_2}+2\psi_{t_1}+(\phi_{t_1})^2+(\psi_{t_0})^2=0,
\end{array}
\end{equation}
where $\phi:=\widetilde{S}(z_{\alpha})$, $\psi: =\widetilde{S}_z(z_{\alpha})$%
.

Equation \eqref{3.8} allows us to derive an analog of the relationships %
\eqref{3.9}-\eqref{3.14} for the logarithmic times. Indeed, by substituting %
\eqref{3.18} into \eqref{3.9} one obtains 
\begin{equation}  \label{3.23}
\begin{gathered}
e^{-\mathcal{D}S(z)}-e^{-\mathcal{D}S(z_0)}+\frac{1}{z_1-z_{\alpha}}
e^{(\mathcal{D}+\mathcal{D}(z_1))S(z)}\cdot\\\\
\cdot\Big[e^{-\mathcal{D}(z_1)S(z)}-e^{-\mathcal{D}(z_1)S(z_0)} \Big]=0,
\end{gathered}
\end{equation}
where we denote $\mathcal{D}\equiv \frac{\partial}{\partial t_0}$.

Similar to the case \eqref{3.9}, equation \eqref{3.23} is the generating
equation for the hierarchy of equations \eqref{3.21}. They are obtained by
the expansion of the left-hand side of \eqref{3.23} in Taylor series in $z_1$%
. Then, by evaluating \eqref{3.23} at $z_0=z_1$, one gets 
\begin{equation}  \label{3.24}
e^{-\mathcal{D}(S(z)-\widetilde{S}(z_{\alpha}))}-e^{-\mathcal{D}(S(z_1)-%
\widetilde{S}(z_{\alpha}))} +\frac{1}{z_1-z_{\alpha}}e^{-\mathcal{D}%
(z_1)(S(z)-\widetilde{S}(z_{\alpha}))}=0,
\end{equation}
where $z\in\ensuremath{\mathbb{C}},z_1\in\tilde{G}$, which implies %
\eqref{3.12}. Evaluating \eqref{3.24} at $z=z_2\in\tilde{G}$ and using %
\eqref{3.4} and \eqref{3.12}, one readily obtains the addition formulas for $%
F$: 
\begin{equation}  \label{3.25}
\frac{1}{z_2-z_{\alpha}}e^{\mathcal{D}\mathcal{D}(z_2)F} -\frac{1}{%
z_1-z_{\alpha}}e^{\mathcal{D}\mathcal{D}(z_1)F}+ \frac{z_2-z_1}{%
(z_1-z_{\alpha})(z_2-z_{\alpha})} e^{\mathcal{D}(z_1)\mathcal{D}(z_2)F}=0,\;
\end{equation}
where $z_1,z_2\in\tilde{G}$. Formulae \eqref{3.21},\eqref{3.22}-\eqref{3.25}
incorporate those of the d(KP-mKP) hierarchy. It is easy to see that %
\eqref{3.25} implies the dKP addition formula \eqref{3.14}. It is
straightforward to check that the function $S$, which obeys \eqref{3.21},
solves also equations \eqref{3.6} and, in particular, equations \eqref{3.18}
imply \eqref{3.5}. Thus, it means that the logarithmic source is fundamental
for the description of deformations of the problem \eqref{2.1}.

Finally we consider variations of the positions $z_{\alpha}$ where the
source $\rho$ is concentrated. For the pure logarithmic source $%
S_0=t_0\ln(z-z_{\alpha})$ and 
\begin{equation}  \label{3.26}
\frac{\partial S}{\partial z_{\alpha}}=-\frac{t_0}{z-z_{\alpha}}+ \frac{%
\partial \widetilde{S}}{\partial z_{\alpha}}.
\end{equation}
Using \eqref{3.3}, one deduces that 
\begin{equation}
\frac{\partial S(z)}{\partial z_{\alpha}}+Ae^{-S_{t_0}(z)}+B=0,
\end{equation}
where $A=t_0 \exp{\widetilde{S}_{t_0}(z_{\alpha})}$ and $B=-\frac{\partial S%
}{\partial z_{\alpha}}(z_0)-A\exp{(-S_{t_0}(z_0))},\;\; (z_0\in%
\ensuremath{\mathbb{C}},z_0\neq z_{\alpha})$.

In the generic case \eqref{3.1} 
\begin{equation}  \label{3.28}
\frac{\partial S}{\partial z_{\alpha}}=-\frac{t_0}{z-z_{\alpha}}+
\sum_{n\geq 1} \frac{nt_n}{(z-z_{\alpha})^{n+1}}+\frac{\partial \widetilde{S}%
}{\partial z_{\alpha}}.
\end{equation}

In virtue of \eqref{3.2} one has 
\begin{equation}  \label{3.29}
\triangle S= -\frac{t_0}{z-z_{\alpha}}+ \sum_{n\geq 1} \frac{nt_n}{%
(z-z_{\alpha})^{n+1}}+\triangle \widetilde{S},
\end{equation}
where 
\begin{equation}  \label{3.30}
\triangle\equiv -t_0\frac{\partial}{\partial t_1}+ \sum_{n\geq 1} nt_n\frac{%
\partial}{\partial t_{n+1}}.
\end{equation}
Using \eqref{3.28} and \eqref{3.29}, one concludes that 
\begin{equation}  \label{3.31}
\frac{\partial S}{\partial z_{\alpha}}-\triangle S+E=0,
\end{equation}
where $E=-\frac{\partial S(z_0)}{\partial z_{\alpha}}+\triangle S(z_0),\;\;
z_0\in\ensuremath{\mathbb{C}},z_0\neq z_{\alpha}$.

In particular, in the gauge $S(z_{0})=0$, we have $E=0$ and 
\begin{equation}
\frac{\partial S(z)}{\partial z_{\alpha }}=\triangle S(z).  \label{3.32}
\end{equation}
This relation shows a connection of the variations of $z_{\alpha }$ with non
isospectral deformations for the dKP hierarchy.

\textbf{Remark 3.2 }Equations (34), (36) and the dKP hierarchy are invariant
under the scale transformations

\begin{eqnarray*}
t_{n} &\longrightarrow &t_{n}^{\prime }=\lambda t_{n},\quad n=1,2,...,\quad
-\infty <\lambda <\infty \\
F(t) &\rightarrow &F^{\prime }(t^{\prime })=\lambda ^{2}F(t), \\
S(t) &\rightarrow &S^{\prime }(t^{\prime })=\lambda S(t), \\
u(t) &\rightarrow &u^{\prime }(t^{\prime })=u(t).
\end{eqnarray*}

Variation of the function F due to the infinitesimal variation of the times $%
t_{n}$ ( $\delta t_{n}=\varepsilon t_{n})$ is equal to $\delta F=\varepsilon
\sum_{n\geq 1}t_{n}\frac{\partial F}{\partial t_{n}}.$

\textbf{Remark 3.3} \ Let us introduce the complex Dirichlet type integral

\[
D_{\widetilde{G}}=\frac{1}{2\pi i}\int \int_{\widetilde{G}}(\partial S\wedge 
\overline{\partial }S-\partial S_{0}\wedge \overline{\partial }S_{0}).
\]

For the d(KP-mKP) hierarchy one gets, using (20) with $t_{0}=0$, the
expansion $\widetilde{S}=\sum_{n\geq 0}(z-z_{\alpha })^{n}\widetilde{S}_{n}$
\ as $z\rightarrow z_{\alpha }$ and the formula (32):

\[
D_{\widetilde{G}}=-\sum_{n\geq 1}nt_{n}\widetilde{S}_{n}=\sum_{n\geq 1}t_{n}%
\frac{\partial F}{\partial t_{n}}. 
\]

Thus, $\delta F=\varepsilon D_{\widetilde{G}}.$ An interrelation between the
Dirichlet type integrals and $\tau $-functions for dispersionless
hierarchies will be discussed in a separate paper.

\section{Two-points source. Dispersionless Laplace and 2DTL hierarchies}

The case of a source $\rho(h)$ concentrated in two different points $%
z_{\alpha}$ and $z_{\beta}$ is rather rich. The domain $\widetilde{G}$ can
be a disconnected set ($\widetilde{G}=\mathcal{D}_{\alpha}\bigcup\mathcal{D}%
_{\beta}$), where $\mathcal{D}_{\alpha}$ and $\mathcal{D}_{\beta}$ are two
disks centered at $z_{\alpha}$ and $z_{\beta}$ respectively. Several types
of different reduced hierarchies can arise.

The function $S_0$ has the form 
\begin{equation}  \label{4.1}
S_0=t_{0\alpha}\ln(z-z_{\alpha})+\sum_{n=1}^{\infty}\frac{x_n}{%
(z-z_{\alpha})^n}+ t_{0\beta}\ln(z-z_{\beta})+\sum_{n=1}^{\infty}\frac{y_n}{%
(z-z_{\beta})^n}.
\end{equation}
Let us first consider variations of the variables $x_n$ and $y_n$. Similar
to the one-point case \eqref{3.6} one obtains two hierarchies of equations 
\begin{equation}  \label{4.2}
S_{x_n}=\sum_{k=0}^n U_k(x,y)(S_{x_1})^k,
\end{equation}
\begin{equation}  \label{4.3}
S_{y_n}=\sum_{k=0}^n V_k(x,y)(S_{y_1})^k,\quad n\geq 1,
\end{equation}
where $U_k$ and $V_k$ are certain functions. Each of these hierarchies gives
rise to the dKP-mKP hierarchy in the variables $x_n$ and $y_n$,
respectively. Interconnection between both hierarchies is provided by the
equation 
\begin{equation}  \label{4.4}
S_{x_1}S_{y_1}+aS_{x_1}+bS_{y_1}+c=0,
\end{equation}
where 
\begin{equation}  \label{4.5}
a=-S_{y_1}(z_{\alpha}),\quad b=-S_{x_1}(z_{\beta}),
\end{equation}
and 
\begin{equation}  \label{4.6}
c=-S_{x_1}(z_0)S_{y_1}(z_0)-aS_{x_1}(z_0)-bS_{y_1}(z_0).
\end{equation}
Here $z_0\in\ensuremath{\mathbb{C}},\;z_0\neq z_{\alpha},z_{\beta}$.

Equation \eqref{4.4} can be treated as the quasi-classical limit of the
well-known Laplace equation 
\begin{equation}  \label{4.7}
\psi_{\xi\eta}+A\psi_{\xi}+B\psi_{\eta}+c\psi=0.
\end{equation}
Indeed, by introducing slow variables $x_1:=\epsilon\xi,\;y_1:=\epsilon\eta$%
, considering the quasiclassical wave-function limit $\psi=\exp(S/\epsilon)$
and proceeding to the limit $\epsilon\rightarrow 0$, provided that $A(\frac{%
x_1}{\epsilon})\rightarrow a(x_1,y_1),\;B(\frac{x_1}{\epsilon})\rightarrow
b(x_1,y_1),\;C(\frac{x_1}{\epsilon})\rightarrow c(x_1,y_1)$, one converts %
\eqref{4.7} into \eqref{4.4}.

So one can refer to the hierarchy described by equations \eqref{4.2}-%
\eqref{4.4} as the dispersionless Laplace hierarchy. It contains different
interesting particular cases. In the gauge $S(z_{0})=0$, the lowest member
of this hierarchy is given by the equations 
\begin{equation}
\begin{gathered} S_{x_1}S_{y_1}+aS_{x_1}+bS_{y_1}+c=0,\\\\
S_{x_2}=(S_{x_1})^2+U_1S_{x_1},\quad S_{y_2}=(S_{y_1})^2+V_1S_{y_1},
\end{gathered}  \label{4.8}
\end{equation}
where $U_{1}=-2\widetilde{S}_{x_{1}}(z_{\alpha }),\;V_{1}=-2\widetilde{S}%
_{y_{1}}(z_{\beta })$. Evaluating equations \eqref{4.8} at $z=z_{\alpha
},z_{\beta }$, one gets the following equations 
\begin{equation}
\begin{gathered}
U_{x_2}=\frac{2}{z_{\beta}-z_{\alpha}}(U_{x_1}-V_{x_1})+(U_{x_1})^2-
2U_{x_1}V_{x_1},\\ V_{x_2}+(V_{x_1})^2=2\phi_{x_1},\\
U_{x_1}V_{y_1}-\frac{1}{z_{\alpha}-z_{\beta}}V_{y_1}+
\frac{1}{z_{\alpha}-z_{\beta}}U_{x_1}+\phi_{y_1}=0, \end{gathered}
\label{4.10}
\end{equation}

\begin{eqnarray*}
V_{y_{2}} &=&\frac{2}{(z_{\alpha }-z_{\beta })}%
(V_{y_{1}}-U_{y_{1}})+(V_{y_{1}})^{2}-2V_{y_{1}}U_{y_{1}}, \\
U_{y_{2}}+(U_{y_{1}})^{2} &=&2\widetilde{\varphi }_{y_{1}}^{~}, \\
\widetilde{\varphi }_{x_{1}} &=&\varphi _{y_{1}}
\end{eqnarray*}

where $\ \ U=\widetilde{S}(z_{\beta }),\;V=\widetilde{S}(z_{\alpha }),\;\phi
=\frac{\partial \widetilde{S}}{\partial z}\mid _{z=z_{\alpha }},\;\widetilde{%
\phi }=\frac{\partial \widetilde{S}}{\partial z}\mid _{z=z_{\beta }}.$

Equations \eqref{4.8} and \eqref{4.10} are the lowest members of the
dispersionless Laplace hierarchy. In terms of the variable $t:=x_{2}+y_{2}$,
equations \eqref{4.10} are equivalent to the dispersionless limit of the
system of equations considered in \cite{38} (equation (9.6)) and in \cite{39}%
. The equations discussed in \cite{39} are the equations for the
Davey-Stewartson wave function. One more case of interesting application of
the dispersionless Laplace transform corresponds to the constraints $%
S(z_{\alpha })=S(z_{\beta })=const$. This constraint is compatible with
equations \eqref{4.2} and \eqref{4.3} for odd $n$. The lowest member of such
a hierarchy is given by the equations 
\begin{equation}
\begin{gathered} S_{x_1}S_{y_1}+q=0,\\ S_{x_3}=(S_{x_1})^3+U_1S_{x_1},\\
S_{y_3}=(S_{y_1})^3+V_1S_{x_1}, \end{gathered}  \label{4.12}
\end{equation}
where $U_{1}=3\partial _{y_{1}}^{-1}q_{x_{1}},\;V_{1}=3\partial
_{x_{1}}^{-1}q_{y_{1}}$. The corresponding equations for the potential $q$
are 
\begin{equation}
q_{x_{3}}=3(q(\partial _{y_{1}}^{-1}q_{x_{1}}))_{x_{1}},\quad
q_{y_{3}}=3(q(\partial _{x_{1}}^{-1}q_{y_{1}}))_{y_{1}}.  \label{4.13}
\end{equation}
In terms of $\partial _{t}=\partial _{x_{3}}+\partial _{y_{3}}$ one has 
\begin{equation}
q_{t}=3(q(\partial _{y_{1}}^{-1}q_{x_{1}}))_{x_{1}}+3(q(\partial
_{x_{1}}^{-1}q_{y_{1}}))_{y_{1}}.  \label{4.14}
\end{equation}

Equation \eqref{4.14} is the dispersionless limit of the
Nizhnik-Veselov-Novikov equation \cite{40}-\cite{41} 
\[
q_{t}=q_{x_{1}x_{1}x_{1}}+q_{y_{1}y_{1}y_{1}}+3(q(\partial
_{y_{1}}^{-1}q_{x_{1}})_{x_{1}}+3(q(\partial
_{x_{1}}^{-1}q_{y_{1}})_{y_{1}}. 
\]
Notice that the dispersionless limit of the Veselov-Novikov equation was
discussed for the first time in \cite{4}.

Equations \eqref{4.12} have an interesting application in physics. Namely,
the first equation is, in fact, the eiconal equation from the geometric
optics. Indeed, if we denote $z=x_1,\;\bar{z}=y_1,\; z=x+iy$, it reads 
\[
S_x^2+S_y^2=-4q. 
\]

\noindent \textbf{Remark 4.1} For the multipoint sources case, the gauge $%
S(z_0)=const.,\; z_0\neq z_{\alpha},z_{\beta}$ is a symmetric and natural
one.

\noindent \textbf{Remark 4.2} Equation \eqref{4.4} implies that 
\begin{equation}
S_{y_1}=-\frac{aS_{x_1}+c}{S_{x_1}+b}.
\end{equation}
This relation shows that, using only one source $z_{\alpha}$, one can
generate infinitesimal deformations of $S$ which have poles at other points
(in this case at $z= z_{\beta})$. This fact, which was already discussed in
Section 2, means that, in particular, it is convenient to introduce the
corresponding times from the very beginning.

Let us include now the logarithmic times into consideration. Analogously to %
\eqref{3.21}, we have in the gauge $S(z_0)=0$ 
\begin{equation}  \label{4.16}
\begin{gathered} S_{x_n}=\sum_{k=0}^n A_ke^{-kS_{t_{0\alpha}}},\\\\
S_{y_n}=\sum_{k=0}^n B_ke^{-kS_{t_{0\beta}}},\quad n\geq 1, \end{gathered}
\end{equation}
where $A_k$ and $B_k$ are functions depending on the whole set of times. In
addition there is an equation relating the derivatives of $S$ with respect
to the times $t_{0\alpha}$ and $t_{0\beta}$ 
\begin{equation}  \label{4.18}
e^{-S_{t_{0\alpha}}-S_{t_{0\beta}}}+\widetilde{A}e^{-S_{t_{0\alpha}}}+ 
\widetilde{B}e^{-S_{t_{0\beta}}}+\widetilde{C}=0,
\end{equation}
where 
\begin{equation}  \label{4.19}
\begin{gathered}
\widetilde{A}=\frac{1}{z_{\beta}-z_{\alpha}}e^{-S_{t_{0\beta}}(z_{\alpha})},%
\\\\
\widetilde{B}=\frac{1}{z_{\alpha}-z_{\beta}}e^{-S_{t_{0\alpha}}(z_{\beta})},%
\quad \widetilde{C}=-1-\widetilde{A}-\widetilde{B}. \end{gathered}
\end{equation}

Equations \eqref{4.16}-\eqref{4.18} describe the general two-points
hierarchy of the infinitesimal deformations. It contains a well-known
dispersionless integrable hierarchy. Indeed, let us introduce new variables $%
T$ and $t$ such that $t_{0\alpha }=T+t,\;t_{0\alpha }=T-t$. Then one has 
\begin{equation}
e^{\pm S_{t}}=\Big(\frac{z-z_{\alpha }}{z-z_{\beta }}\Big)^{\pm 1}e^{\pm 
\widetilde{S}_{t}}.  \label{4.20}
\end{equation}
Using \eqref{4.20}, one obtains 
\begin{equation}
S_{x_{n}}=\sum_{k=0}^{n}U_{k}e^{-kS_{t}},\quad
S_{y_{n}}=\sum_{k=0}^{n}V_{k}e^{kS_{t}},\quad n\geq 1,  \label{4.21}
\end{equation}
for certain functions $U_{k}$ and $V_{k}$. The lowest member of the
hierarchy \eqref{4.21} is 
\begin{equation}
S_{x_{1}}=Ue^{-S_{t}}-U,\quad S_{y_{1}}=Ve^{S_{t}}-V,  \label{4.23}
\end{equation}
where 
\[
U=\frac{1}{z_{\alpha }-z_{\beta }}e^{\widetilde{S}_{t}(z_{\alpha })},\quad V=%
\frac{1}{z_{\beta }-z_{\alpha }}e^{-\widetilde{S}_{t}(z_{\beta })} 
\]
Evaluating the left-hand sides of the equations \eqref{4.23} at $z=z_{\beta
} $ and $z_{\alpha }$, one gets 
\begin{equation}
\phi _{x_{1}}=\frac{1}{z_{\alpha }-z_{\beta }}\Big(1-e^{\psi _{t}}\Big)%
,\quad \psi _{y_{1}}=\frac{1}{z_{\beta }-z_{\alpha }}\Big(1-e^{-\phi _{t}}%
\Big),  \label{4.25}
\end{equation}
where $\phi =\widetilde{S}(z_{\beta }),\;\psi =\widetilde{S}(z_{\alpha })$,
or 
\begin{equation}  \label{4.26}
U_{y_{1}}+UV_{t}=0,\quad V_{x_{1}}-VU_{t}=0.
\end{equation}
The system \eqref{4.25} implies 
\begin{equation}
\theta _{x_{1}y_{1}}+\Big(e^{\theta }\Big)_{tt}=0,  \label{4.27}
\end{equation}
where 
\begin{equation}
\theta =\ln \Big[-\frac{1}{(z_{\alpha }-z_{\beta })^{2}}e^{\psi _{t}-\phi
_{t}}\Big]=\lg (UV).
\end{equation}
Equation \eqref{4.27} is the well-known dispersionless 2DTL equation (see 
\cite{11}). Its associated system of Hamilton-Jacobi equations is provided
by \eqref{4.21} \cite{11},\cite{21}.

Note that equations \eqref{4.23} imply the following equation for $S$ 
\begin{equation}  \label{4.29}
S_{x_1y_1}+S_{x_1}S_{y_1}\Big(\frac{1}{e^{S_t}-1}\Big)_t=0.
\end{equation}

Now we will derive generating and addition formulae for the dispersionless
2DTL hierarchy in the gauge $S(z_0)=0$. To this end we introduce the
operators 
\begin{equation}  \label{4.30}
\mathcal{D}_{\alpha}(z):=\sum_{n=1}^{\infty}\frac{1}{n}(z-z_{\alpha})^n 
\frac{\partial}{\partial x_n},\quad \mathcal{D}_{\beta}(z):=\sum_{n=1}^{%
\infty}\frac{1}{n}(z-z_{\beta})^n \frac{\partial}{\partial y_n}.
\end{equation}
One has 
\begin{equation}  \label{4.31}
\begin{gathered}
e^{-\mathcal{D}_{\alpha}(z_1)S(z)}=\frac{z-z_1}{z-z_{\alpha}}
e^{-\mathcal{D}_{\alpha}(z_1)\widetilde{S}(z)},\\\\
e^{-\mathcal{D}_{\beta}(z_2)S(z)}=\frac{z-z_2}{z-z_{\beta}}
e^{-\mathcal{D}_{\beta}(z_2)\widetilde{S}(z)}, \end{gathered}
\end{equation}
where $z\in\ensuremath{\mathbb{C}},\;z_1,z_2\in\widetilde{G}$. Using %
\eqref{4.31} one obtains for $p_{\alpha}:=\frac{\partial S}{\partial x_1}$
and $p_{\beta}:=\frac{\partial S}{\partial y_1}$ 
\begin{equation}  \label{4.33}
\begin{gathered} p_{\alpha}(z)-p_{\alpha}(z_1)+\frac{1}{z_1-z_{\alpha}}
e^{-\mathcal{D}_{\alpha}(z_1)(S(z)-\widetilde{S}(z_{\alpha}))}=0,\;
z\in\ensuremath{\mathbb{C}},\;z_1\in\widetilde{G},\\\\
p_{\beta}(z)-p_{\beta}(z_2)+\frac{1}{z_2-z_{\beta}}
e^{-\mathcal{D}_{\beta}(z_2)(S(z)-\widetilde{S}(z_{\beta}))}=0,\;
z\in\ensuremath{\mathbb{C}},\;z_2\in\widetilde{G}. \end{gathered}
\end{equation}

By proceeding as in the one-point case, one finds that equations \eqref{4.33}
imply that 
\begin{equation}  \label{4.35}
\widetilde{S}(z)=\widetilde{S}(z_{\alpha})- \mathcal{D}_{\alpha}(z)F_{%
\alpha},\quad \widetilde{S}(z)=\widetilde{S}(z_{\beta})- \mathcal{D}%
_{\beta}(z)F_{\beta},\; z\in\widetilde{G},
\end{equation}
where $F_{\alpha}$ and $F_{\beta}$ are functions depending on the times
only, which satisfy addition formulae of the form \eqref{3.14}.

On the other hand, using \eqref{4.20} one gets the identities 
\begin{equation}  \label{4.36}
\begin{gathered} p_{\alpha}(z)-p_{\alpha}(z_{\beta})=\frac{1}{z_{\alpha}
-z_{\beta}}e^{-D(S(z)-\widetilde{S}(z_{\alpha}))},\\\\
p_{\beta}(z)-p_{\beta}(z_{\alpha})=\frac{1}{z_{\beta}
-z_{\alpha}}e^{-D(S(z)-\widetilde{S}(z_{\beta}))}, \end{gathered}
\end{equation}
where $z\in\ensuremath{\mathbb{C}}$ and we denote $D:=\frac{\partial}{%
\partial t}$. By substituting \eqref{4.36} into \eqref{4.33}, we obtain 
\begin{equation}  \label{4.38}
\begin{aligned} \frac{1}{z_{\alpha}
-z_{\beta}}&e^{-D(S(z)-\widetilde{S}(z_{\alpha}))} -\frac{1}{z_{\alpha}
-z_{\beta}}e^{-D(S(z_1)-\widetilde{S}(z_{\alpha}))}\\
&+\frac{1}{z_1-z_{\alpha}}e^{-\mathcal{D}_{\alpha}(z_1)(S(z)-%
\widetilde{S}(z_{\alpha}))}=0, \quad
z\in\ensuremath{\mathbb{C}},\;z_1\in\widetilde{G}, \end{aligned}
\end{equation}
and 
\begin{equation}  \label{4.39}
\begin{aligned} \frac{1}{z_{\beta}
-z_{\alpha}}&e^{-D(S(z)-\widetilde{S}(z_{\beta}))} -\frac{1}{z_{\beta}
-z_{\alpha}}e^{-D(S(z_2)-\widetilde{S}(z_{\beta}))} \\
&+\frac{1}{z_2-z_{\beta}}e^{-\mathcal{D}_{\beta}(z_2)(S(z)-\widetilde{S}(z_{%
\beta}))}=0,\quad \quad z\in\ensuremath{\mathbb{C}},\;z_2\in\widetilde{G}.
\end{aligned}
\end{equation}
These identities are generating equations for \eqref{4.21}. Indeed,
expanding their left-hand sides in Taylor series in $z_1$ and $z_2$, one
gets \eqref{4.21}. Furthermore, evaluating \eqref{4.38} at $z=z_{\beta}$ and 
$z=z_{\alpha}$, one deduces that $\widetilde{S}(z_{\alpha})-\widetilde{S}%
(z_{\beta})=DF_{\alpha}=DF_{\beta}$ and so on. Hence 
\begin{equation}  \label{4.49}
F_{\alpha}=F_{\beta}=F=\ln\tau_{\mbox{d2DTL}}.
\end{equation}
Thus there is only one $\tau$-function for the d2DTL hierarchy. This feature
is in agreement with earlier results \cite{11}. Considering \eqref{4.38} at $%
z=\tilde{z}_1\in\widetilde{G}$ and $z=\tilde{z}_2\in\widetilde{G}$, and
using \eqref{4.39}, one shows that the function $F$ satisfies 
\begin{equation}  \label{4.41}
\begin{aligned}
(\tilde{z}_1-z_{\beta})&(z_1-z_{\alpha})e^{D\mathcal{D}_{\alpha}(%
\tilde{z}_1)F}-
(\tilde{z}_1-z_{\beta})(\tilde{z}_1-z_{\alpha})e^{D\mathcal{D}_{%
\alpha}(z_1)F}+\\ &+(z_{\alpha}-z_{\beta})(\tilde{z}_1-z_1)e^{
\mathcal{D}_{\alpha}(z_1)\mathcal{D}_{\alpha}(\tilde{z}_1)F}=0,\\
\end{aligned}
\end{equation}
and 
\begin{equation}  \label{4.42}
\begin{aligned}
(\tilde{z}_2-z_{\alpha})&(z_2-z_{\beta})e^{-D\mathcal{D}_{\beta}(%
\tilde{z}_2)F}-
(\tilde{z}_2-z_{\alpha})(\tilde{z}_2-z_{\beta})e^{-D\mathcal{D}_{%
\alpha}(z_2)F}+\\ &+(z_{\beta}-z_{\alpha})(\tilde{z}_2-z_2)e^{
\mathcal{D}_{\beta}(z_2)\mathcal{D}_{\alpha}(\tilde{z}_2)F}=0.\\
\end{aligned}
\end{equation}
These identities imply, in particular, the addition formulae for the dKP
hierarchy.

Then, evaluating \eqref{4.38} at $z=z_2$, using the relations $\widetilde{S}%
(z_2)=\widetilde{S}(z_{\beta})-\mathcal{D}_{\beta}(z_2)F$ and $\widetilde{S}%
(z_{\alpha})-\widetilde{S}(z_{\beta})=DF$, one obtains 
\begin{equation}  \label{4.43}
\begin{aligned} 1&+\frac{(z_{\alpha}-z_{\beta})(z_1-z_{\alpha})}
{(z_1-z_{\beta})(z_2-z_{\alpha})}e^{\mathcal{D}_{\alpha}(z_1)
\mathcal{D}_{\beta}(z_2)F}\\ &-\frac{(z_1-z_{\alpha})(z_2-z_{\beta})}
{(z_1-z_{\beta})(z_2-z_{\alpha})}e^{[\mathcal{D}_{\beta}(z_2)-
\mathcal{D}_{\alpha}(z_1)+D]DF}=0. \end{aligned}
\end{equation}
Evaluating equation \eqref{4.39} at $z=z_1$ and taking into account that $%
\widetilde{S}(z_1)=\widetilde{S}(z_{\alpha})-\mathcal{D}_{\alpha}(z_1)F$ ,
one gets the same equation \eqref{4.43}.

Equations \eqref{4.41}, \eqref{4.42} and \eqref{4.43} form a complete set of
addition formulae for the d2DTL hierarchy which completely define \cite{11}, 
\cite{18} its associated $\tau$ function.

Similarly to the one-point case, one can derive a generating equation for
the hierarchy of equations for $S$(see \cite{37})

Finally, let us include into consideration the variations of positions $%
z_{\alpha}$ and $z_{\beta}$ of singularities. The simplest case is given by
the one logarithmic source 
\begin{equation}  \label{4.44}
S_0=t\ln\frac{z-z_{\alpha}}{z-z_{\beta}}.
\end{equation}
Since 
\begin{equation}  \label{4.45}
S_{z_{\alpha}}=-\frac{t}{z-z_{\alpha}}+\frac{\partial \widetilde{S}}{%
\partial z_{\alpha}},\quad S_{z_{\beta}}=\frac{t}{z-z_{\beta}}+\frac{%
\partial \widetilde{S}}{\partial z_{\beta}},
\end{equation}
one obtains the following equations (we use the gauge $S(z_0)=0$) 
\begin{equation}  \label{4.46}
S_{z_{\alpha}}=Ue^{-S_t}-U,\quad S_{z_{\beta}}=Ve^{S_t}-V,
\end{equation}
where 
\begin{equation}  \label{4.47}
U=\frac{t}{z_{\alpha}-z_{\beta}}e^{\widetilde{S}_t(z_{\alpha})},\quad V=%
\frac{t}{z_{\beta}-z_{\alpha}}e^{-\widetilde{S}_t(z_{\beta})}.
\end{equation}

Equations \eqref{4.46} imply that $\theta=\lg(UV)$ obeys 
\begin{equation}  \label{4.48}
\theta_{z_{\alpha}z_{\beta}}+(e^{\theta})_{tt}=0,
\end{equation}
which is again the d2DTL equation \eqref{4.27}, but now the spacial
variables are the positions of the logarithmic singularities. In particular
for $z_{\alpha}=w,\; z_{\beta}=\bar{w}$ one gets the elliptic version of the
d2DTL hierarchy. Despite of the fact that \eqref{4.48} coincides with %
\eqref{4.27} their corresponding dressing procedures are different. Indeed,
for the simplest choice $\widetilde{S}(z_{\alpha}):=\psi=0,\; \widetilde{S}%
(z_{\beta}):=\phi=0,$ the formula \eqref{4.29} gives the trivial solution $%
\theta=const.$ of \eqref{4.27}, while for \eqref{4.48} one gets the
nontrivial solution 
\[
\theta=\lg(UV)=2\lg\Big(\frac{t}{z_{\beta}-z_{\alpha}}\Big). 
\]
For the whole d2DTL hierarchy one has two relations of the type \eqref{3.31}.

\section{Multi-point sources}

For the general $N$ points-like source 
\begin{equation}  \label{5.1}
S_0=\sum_{\alpha=1}^N t_{0\alpha}\lg(z-z_{\alpha})+\sum_{\alpha=1}^N
\sum_{n=1}^{\infty}\frac{t_{n\alpha}}{(z-z_{\alpha})^n},
\end{equation}
one has $N$ hierarchies of equations 
\begin{equation}  \label{5.2}
S_{t_{n\alpha}}=\sum_{k=0}^n U_{\alpha k}e^{-kS_{t_{0\alpha}}},\quad
\alpha=1,\ldots,N,\; n\geq 1.
\end{equation}
These families of equations are related by equations of the type \eqref{4.18}
\begin{equation}  \label{5.3}
e^{-S_{t_{0\alpha}}-S_{t_{0\beta}}}+A_{\alpha\beta}e^{-S_{t_{0\alpha}}}+
A_{\beta\alpha}e^{-S_{t_{0\beta}}}+B_{\alpha\beta}=0,
\end{equation}
where 
\begin{equation}
A_{\alpha\beta}=e^{-S_{t_{0\beta} }(z_{\alpha})},\quad
B_{\alpha\beta}=-(1+A_{\alpha\beta}+A_{\beta\alpha}).
\end{equation}
The hierarchy of equations \eqref{5.2}, \eqref{5.3} is nothing but the
universal Whitham hierarchy introduced in \cite{8}. It contains a number of
interesting subhierarchies. Obviously, there are $N$ d(KP-mKP) hierarchies
associated with each of the singularity points $z_{\alpha} \;
(\alpha=1,\ldots,N)$. Then there are d2DTL hierarchies corresponding to each
pair $(z_{\alpha},z_{\beta})$ of the singularity points. One can show that
there is only one function $F$ for \eqref{5.2},\eqref{5.3} \cite{8} .To
demonstrate this property it is enough to apply the same argumentation as
that used in the previous section to identify each pair $F_{\alpha}$ and $%
F_{\beta}$.

We will discuss now some of the subhierarchies of \eqref{5.2},\eqref{5.3}
with more that two singularity points. Firstly, by restricting ourselves to
the variables $t_{1\alpha}\;(\alpha=1,\ldots,N)$, we obtain the following
set of equations (in the gauge $S(z_0)=0$) 
\begin{equation}  \label{5.5}
S_{\xi_{\alpha}}S_{\xi_{\beta}}-\varphi_{\xi_{\beta}}^{\alpha}S_{\xi_{%
\alpha}}- \varphi_{\xi_{\alpha}}^{\beta}S_{\xi_{\beta}}=0,\quad
\alpha\neq\beta,\; \alpha,\beta=1,\ldots,N,
\end{equation}
where $\xi_{\alpha}:=t_{1\alpha}$ and $\varphi^{\alpha}:=S(z_{\alpha})$. By
evaluating the left-hand side of \eqref{5.5} at the points $z=z_{\gamma}\;
(\gamma\neq\alpha,\beta)$ one gets the bilinear system of equations 
\begin{equation}  \label{5.6}
\varphi_{\xi_{\alpha}}^{\gamma}\varphi_{\xi_{\beta}}^{\gamma}
-\varphi_{\xi_{\beta}}^{\alpha}\varphi_{\xi_{\alpha}}^{\gamma}
-\varphi_{\xi_{\alpha}}^{\beta}\varphi_{\xi_{\beta}}^{\gamma}=0
\end{equation}
where $\alpha,\beta$ and $\gamma$ are different. Equivalently 
\begin{equation}  \label{5.7}
\frac{\varphi_{\xi_{\beta}}^{\alpha}}{\varphi_{\xi_{\beta}}^{\gamma}}+ \frac{%
\varphi_{\xi_{\alpha}}^{\beta}}{\varphi_{\xi_{\alpha}}^{\gamma}}=1.
\end{equation}
Equations \eqref{5.5} and \eqref{5.6} can be treated as the dispersionless
limits of the systems 
\begin{equation}  \label{5.8}
\psi_{u_{\alpha}u_{\beta}}=(\lg
H^{\alpha})_{u_{\beta}}\psi_{u_{\alpha}}+(\lg
H^{\beta})_{u_{\alpha}}\psi_{u_{\beta}},
\end{equation}
\begin{equation}  \label{5.9}
H_{u_{\alpha}u_{\beta}}^{\gamma}= (\lg
H^{\alpha})_{u_{\beta}}H_{u_{\alpha}}^{\gamma}+(\lg
H^{\beta})_{u_{\alpha}}H_{u_{\beta}}^{\gamma},
\end{equation}
where $\alpha,\beta$ and $\gamma$ are different. The equations \eqref{5.9}
form the Darboux system describing the $N$-conjugate systems of surfaces in
Euclidean space, and \eqref{5.8} are the corresponding equations for the
position vector \cite{42}. In this case the slow variables are $%
\xi_{\alpha}=\epsilon u_{\alpha}$ and the quasiclassical limit is
implemented by the expressions $\psi=\exp{\frac{S}{\epsilon}}%
,\;H^{\alpha}=\exp{\frac{\varphi^{\alpha}}{\epsilon}}$. Thus, one shows that %
\eqref{5.8} anbd \eqref{5.9} represent the dispersionless limit of the
Darboux system and its associated linear system (see also \cite{37}).

Our second example is determined by an $S_0$ function of the form 
\begin{equation}  \label{5.10}
S_0=\frac{x}{z-z_0}+y\sum_{n=1}^N \frac{a_n}{z-z_n}+t\sum_{m=1}^M \frac{b_m}{%
z-\tilde{z}_m},
\end{equation}
where $a_n,\; b_m$ are arbitrary complex parameters and the sets $\{z_n\}$
and $\{\tilde{z}_m\}$ are disjoint. In this case 
\begin{equation}  \label{5.11}
\begin{gathered} p:=S_x=\frac{1}{z-z_0}+\widetilde{S}_x,\\ S_y=\sum_{n=1}^N
\frac{a_n}{z-z_n}+\widetilde{S}_y,\\ S_t=\sum_{m=1}^M
\frac{b_m}{z-\tilde{z}_m}+\widetilde{S}_t. \end{gathered}
\end{equation}
From these expressions the following Hamilton-Jacobi type equations follow 
\begin{equation}  \label{5.12}
S_y=V_0+\sum_{n=1}^N \frac{V_n}{S_x-U_n},\quad S_t=\widetilde{V}%
_0+\sum_{m=1}^M \frac{\widetilde{V}_m}{S_x-\widetilde{U}_m},
\end{equation}
where 
\begin{equation}  \label{5.13}
U_n=S_x(z_n,x,y,t),\quad \widetilde{U}_m=S_x(\tilde{z}_m,x,y,t),
\end{equation}
and $V_0,\;V_n,\;\widetilde{V}_0,\;\widetilde{V}_m$ are functions of $%
(x,y,t) $. The type of equations \eqref{5.12} as well as its associated
integrable systems have been considered in \cite{10}.

Another interesting subhierarchy corresponds to the choice 
\begin{equation}  \label{5.14}
S_0=\frac{x}{z-z_0}+\frac{t}{(z-z_0)^2}+y\sum_{n=1}^N \frac{a_n}{z-z_n}.
\end{equation}
In this case one has the following equations 
\begin{equation}  \label{5.15}
S_y=V_0+\sum_{n=1}^N \frac{V_n}{S_x-U_n},\quad S_t=S_x^2+U_0,
\end{equation}
where $U_n:=S_x(z_n,x,y,t)$. The associated integrable system 
\begin{equation}  \label{5.16}
\begin{gathered} U_{n,t}+U_{0,x}-(U_n^2)_x=0,\\ V_{n,t}-2(V_nU_n)_x=0,\;\;
1\leq n\leq N,\\ U_{0,y}+2\sum_{i=1}^N V_{n,x}=0, \end{gathered}
\end{equation}
is the generalization of the Benney system to $2+1$ dimensions proposed in 
\cite{8},\cite{10}

Other particular cases of the hierarchy \eqref{5.2},\eqref{5.3} will be
considered elsewhere.

\vspace{0.3cm} \noindent \textbf{Remark 5.1} All of our constructions in
this and previous sections were local ones, with singularities (sources)
located in certain domains of $\ensuremath{\mathbb{C}}$. This means that one
will get the same formulae by considering instead a chart of the Riemann
sphere. Compactness of the Riemann sphere imposes certain constraints of the
form \eqref{5.1} of the sources (singularities). In particular, $%
\sum_{\alpha =1}^{N}t_{0\alpha }=0$ (see e.g.\cite{34}).

\vspace{0.3cm} \noindent \textbf{Acknowledgements} The authors are very
grateful to \emph{the Isaac Newton Institute for Mathematical Sciences} of
Cambridge, where this work has been done and written, for the kind
hospitality. They are also grateful to the organizers of the programme
"Integrable Systems" for the support provided. L. Martinez Alonso wish to
thank the \emph{Fundaci\'{o}n Banco Bilbao Vizcaya Argentaria} for
supporting his stay at Cambridge University as a BBV visiting professor.


\begin{thebibliography}{99}
\bibitem{1}  B. A. Kupershmidt and Yu. I. Manin, Long wave equations with a
free boundary, I , \emph{Funkts. Anal. Prilozh}.\textbf{\ 11}(3): 31 (1977); 
\textbf{II, }\emph{Funkts .Anal. Prilozh}. \textbf{12} (1): 25 (1978); \
D.R. Lebedev and Yu.I. Manin, Conservation laws and Lax representation of
Benney`s long wave equations, \emph{Phys. Lett. }\textbf{74A : }154-156
(1979).

\bibitem{2}  V. E. Zakharov, \ Benney equations and quasi-classical
approximation in the inverse problem method, \emph{Func. Anal. Priloz}. 
\textbf{14}: 89-98 (1980).

\bibitem{3}  P. D. Lax and C. D. Levermore, The small dispersion limit on
the Korteweg-de Vries equation, \emph{Commun. Pure Appl. Math.} \textbf{36}:
253-290, 571-593, 809-830 (1983 ).

\bibitem{4}  I. M. Krichever, Averaging method for two-dimensional
integrable equations,\emph{\ Funkts . Anal. Priloz}. \textbf{22}: 37-52
(1988).

\bibitem{5}  Y. Kodama, A method for solving the dispersionless KP equation
and its exact solutions, \emph{Phys. Lett}. \textbf{129A}: 223-226 (1988); \
Y. Kodama, Solutions of the dispersionless Toda equation, \emph{Phys.Lett}. 
\textbf{147A}: 477-482 (1990).

\bibitem{6}  B. A. Dubrovin and S. P. Novikov, Hydrodynamics of weakly
deformed soliton lattices: differential geometry and Hamiltonian theory, 
\emph{Russian Math. Surveys,} \textbf{44}: 35-124 (1989).

\bibitem{7}  K. Takasaki and T. Takebe,\emph{\ \ }SDIFF(2) KP hierarchy, 
\emph{\ Int. J. Mod. Phys. A}, Vol.\textbf{7}, Suppl.\textbf{1B}: 889-922
(1992).

\bibitem{8}  I. M. Krichever, The $\tau $-function of the universal Whitham
hierarchy, matrix models and topological field theories, \ \emph{Commun.
Pure Appl. Math.} \textbf{47}: 437-475 (1994).

\bibitem{9}  \emph{Singular limits of dispersive waves} (eds. N. M. Ercolani
et al), Nato Adv. Sci. Inst. Ser. B Phys. \textbf{320 }, Plenum, New York
(1994).

\bibitem{10}  V. E. Zakharov, Dispersionless limit of integrable system in
2+1 dimensions, in \cite{9}, pp 165-174 (1994).

\bibitem{11}  K. Takasaki and T. Takebe, Integrable hierarchies and
dispersionless limit, \ \emph{Rev. Math. Phys. }\textbf{7}: 743-818 (1995).

\bibitem{12}  R. Carroll and Y. Kodama, \ Solutions of the dispersionless
Hirota equations, \ \emph{J. Phys. A Math. Gen}. \textbf{28}: 6373-6387
(1995).

\bibitem{13}  S. Jin, C. D. Levermore and D. W. McLaughlin, \ The
semiclassical \ limit of the defocusing NLS hierarchy, \ \emph{Comm. Pure
and Appl. Math.} \textbf{52}: 613-654 (1999).

\bibitem{14}  I. M. Krichever, \ The dispersionless Lax equations and \
topological minimal models, \ \emph{Commun. Math. Phys}. \textbf{143}:
415-429 (1992).

\bibitem{15}  B. A. Dubrovin, \ Hamiltonian formalism of Whitham-type
hierarchies and topological Landau-Ginzburg models, \ \emph{Commun. Math.
Phys.} \textbf{145}: 195-203 (1992); B. A. Dubrovin and Y. Zhang,
Bihamiltonian hierarchies in 2D topological field theory at one-loop
approximation, \ \emph{Commun. Math. Phys}. \textbf{198}: 311-361 (1998); \
B.A.Dubrovin and \ Y.Zhang, \ Normal forms of hierarchies of integrable
PDEs, Frobenious manifolds and Gromov-Witten invariants,
arXiv:math.DG/0108160 v1 (2001).

\bibitem{16}  S. Aoyama and Y. Kodama, Topological \ Landau-Ginzburg theory
with rational potential and the dispersionless KP hierarchy, \ \emph{%
Commun.Math.Phys}. \textbf{182}: 185-219 (1996) ; M. Dunaiski, L.J. Mason
and P. Tod, Einstein-Weyl geometry, the dKP equation and twistor theory, \ 
\emph{J.Geom.Phys.} \textbf{37}: \ 63-93 (2001); L.A. Takhtajan, \ Free
bosons and tau-function for compact Riemann surfaces and smooth Jordan
curves I. Current correlation functions, \emph{Lett.Math.Phys}. \textbf{56}
: 181-228 (2001).

\bibitem{17}  J. Gibbons and S. P. Tsarev, Conformal maps and reductions of
the Benney equations, \ \emph{Phys. Lett.} \textbf{258A}: 263-271 (1999).

\bibitem{18}  P. B. Wiegmann and A. Zabrodin, Conformal maps and integrable
hierarchies, \ \emph{Commun. Math. Phys}. \textbf{213}: 523-538 (2000); M.
Mineev-Weinstein, P. B. Wiegmann and A. Zabrodin, \ Integrable structure of
interface dynamics, \emph{Phys}. \emph{Rev. Lett}. \textbf{84: }5106-5108
(2000) ; \ I.K. Kostov, I.Krichever, M. Mineev-Weinstein, P.B. Wiegmann and
A. Zabrodin, \ $\tau $-function for analytic curves, \ in \emph{\ Random
matrices and their applications , }MSRI Publications, v. \textbf{40, (}%
2001), pp 1-15 ; \textbf{\ }\ A. Marshakov, P. Wiegmann and A. Zabrodin, \ \
Integrable structure of the Dirichlet boundary problem in two dimensions,
arXiv:hep-th/0109048 (2001).

\bibitem{19}  B. G. Konopelchenko, L. Martinez Alonso and O. Ragnisco, \ The 
$\overline{\partial }$- approach to the dispersionless KP hierarchy, \emph{%
J. Phys. A: Math. Gen}. \textbf{34}: 10209-10217 (2001).

\bibitem{20}  B. G. Konopelchenko and L. Martinez Alonso,\emph{\ \ }$%
\overline{\partial }$- equations, integrable deformations of quasiconformal
mappings and Whitham hierarchy, \emph{\ Phys. Lett. }\textbf{286A}: 161-166
(2001).

\bibitem{21}  B. G. Konopelchenko and L. Martinez Alonso, Dispersionless
scalar hierarchies, Whitham hierarchy and the\emph{\ }quasi-classical $\bar{%
\partial}$-method, arXiv: nlin: SI/0105071 (2001) .

\bibitem{22}  L. V. Ahlfors, \emph{Lectures on quasi-conformal mappings}, D.
Van Nostrand C. , Princeton, 1966.

\bibitem{23}  Q. Lehto and K. I. Virtanen, \emph{Quasiconformal mappings in
the plane}, Springer-Verlag, Berlin, 1973.

\bibitem{24}  S. L. Krushkal, \emph{Quasiconformal mappings and Riemann
surfaces}, John Wiley and Sons, New-York, 1979.

\bibitem{25}  R. P. Gilbert and Wen Guo-Chun, \ \ Free boundary problems
occuring in planar fluid dynamics, \emph{Nonlinear Anal.}, \emph{Theory,
Methods, Appl.} \textbf{13}: 285-303 (1989).

\bibitem{26}  M. Boukrouche, \ The quasiconformal mapping methods to solve a
free boundary problem for generalized Hele-Show flows, \emph{Complex
Variables} \textbf{28}: 227-242 (1996).

\bibitem{27}  B. Bojarski, Quasiconformal mappings and general structural
properties of systems of nonlinear equations elliptic in\emph{\ }the\emph{\ }%
sense of Lavrent'ev ,\emph{\ Symposia Mathematica} XVIII: 485-499 (1976).

\bibitem{28}  T. Iwaniec, Quasiconformal mapping problem for general
nonlinear systems of partial differential equations , \emph{Symposia} \emph{%
Mathematica} XVIII: 501-517 (1976).

\bibitem{29}  O. Martio, Partial differential equations and quasiregular
mappings, in \emph{Lecture Notes in Math.} \textbf{Vol 1508}, pp 65-79 (1992)

\bibitem{30}  V. E. Zakharov and S. V. Manakov, \ Construction of
multidimensional nonlinear integrable systems and their solutions, \emph{%
Funkts. Anal. Prilozh}. \textbf{19}: 89-101 (1985).

\bibitem{31}  V. E. Zakharov, On the inverse method, in \emph{Inverse
problems in action} (P. Sabatier, Ed.) p.602, Springer-Verlag, Berlin, 1990.

\bibitem{32}  B. G. Konopelchenko, \emph{Solitons in multidimensions}, World
Scientific, Singapore, 1993.

\bibitem{33}  I. N. Vekua, \emph{Generalized analytic functions}, Pergamon
Press, Oxford, 1962.

\bibitem{34}  A. Hurwitz and R. Courant , \emph{Theory of functions},
Springer-Verlag, Berlin, 1964; G. Springer, \emph{Introduction to Riemann
surfaces}, Chelsea P.C., New York, 1981.

\bibitem{35}  T. Miwa, \ On Hirota`s difference equations, \emph{Proc. Japan
Acad.} \textbf{58} Ser.A : 8-11 (1982).

\bibitem{36}  B. A. Kuperschmidt, The quasiclassical limit of the modified
KP hierarchy, \emph{J. Phys. A: Math. Gen.} \textbf{23}: 876-886 (1990).

\bibitem{37}  L. V. Bogdanov, B. Konopelchenko and L. Martinez Alonso,
Quasi-classical $\bar{\partial}$-dressing method: generating\emph{\ }%
equations for dispersionless integrable hierarchies, arXiv:nlin.SI/0111062
(2001).

\bibitem{38}  B. G. Konopelchenko, Soliton eigenfunction equations: the IST
integrability and some properties, \emph{Rev. Math. Phys}. \textbf{2}:
399-440 (1990).

\bibitem{39}  B. G. Konopelchenko, Nets in $R^{3}$, their integrable
evolutions and the DS hierarchy, \emph{Phys. Lett}. \textbf{183A}: 153-159
(1993).

\bibitem{40}  L. P. Nizhnik, Integration of multidimensional nonlinear
equations by the method of inverse problem, \emph{DAN SSSR} \textbf{254}:
332 (1980).

\bibitem{41}  A. P. Veselov and S. P. Novikov, Finite-zone two-dimensional
potential Schrodinger operators. Explicit formulae and evolution equations, 
\emph{DAN SSSR} \textbf{279}: 20-24 (1984).

\bibitem{42}  G. Darboux, \emph{Lecons sur les systemes orthogonaux et les
coordonnes curvilignes}, Hermann, Paris, 1910.
\end{thebibliography}
\end{document}